\documentclass[sigconf,nonacm,natbib=false,10pt]{acmart}

\usepackage[table]{xcolor}
\usepackage[T1]{fontenc}
\usepackage[utf8]{inputenc}
\usepackage{amsmath}
\usepackage{graphicx}
\usepackage{tikz}
\usepackage{circuitikz}
\usepackage{subcaption}
\usetikzlibrary{positioning, calc, arrows.meta, decorations.pathreplacing, shapes, fit, matrix, shapes.geometric, backgrounds}
\usepackage{etoolbox}
\usepackage{multirow}
\usepackage{txfonts}
\usepackage{listings}
\usepackage{xspace}

\usepackage{hyperref}
\hypersetup{
  colorlinks=true,
    linkcolor=blue,
    filecolor=magenta,
    urlcolor=red,
  }
\usepackage{glossaries}
\glsdisablehyper

\newacronym{engine}{XXX}{AwesomeAlpineFlower}
\newacronym{cep}{CEP}{Complex Event Processing}
\newacronym{cl}{CL}{cache line}
\newacronym{ff}{FF}{flip-flop}
\newacronym{fifo}{FIFO}{first-in-first-out}
\newacronym{hs}{HS}{Home State}
\newacronym{rs}{RS}{Remote State}
\newacronym{vc}{VC}{virtual channel}
\newacronym{thx}{THX}{Thunder-X}
\newacronym{rtg}{RTG}{Remote Tag Store}
\newacronym{dirc}{DirC}{Directory Controller}
\newacronym{cfglut}{CFGLUT5}{Configurable Lookup Table}
\newacronym{ip}{IP}{Integer Program}
\newacronym{ste}{STE}{state transition element}
\newacronym{asic}{ASIC}{Application-Specific Integrated Circuit}
\newacronym{axi}{AXI}{Advanced eXtensible Interconnect}
\newacronym{bdk}{BDK}{Board Development Kit}
\newacronym{ced}{CED}{Complex Event Detection}
\newacronym{clb}{CLB}{configurable logic block}
\newacronym{crc}{CRC}{cyclic redundancy check}
\newacronym[longplural={Deterministic Finite Automata}]{dfa}{DFA}{Deterministic Finite Automaton}
\newacronym{dma}{DMA}{Direct Memory Access}
\newacronym{fpga}{FPGA}{Field Programmable Gate Array}
\newacronym{fsm}{FSM}{Finite-State Machine}
\newacronym{eci}{ECI}{Enzian Coherent Interconnect}
\newacronym{hdl}{HDL}{Hardware Description Language}
\newacronym{hnfa}{hNFA}{homogeneous NFA}
\newacronym{ila}{ILA}{Integrated Logic Analyzer}
\newacronym{ist}{IST}{in-system testing}
\newacronym{lut}{LUT}{Look-Up Table}
\newacronym[longplural={nondeterministic finite automata}]{nfa}{NFA}{nondeterministic finite automaton}
\newacronym{regex}{RegEx}{Regular Expression}
\newacronym{rtl}{RTL}{register transfer language}
\newacronym{rv}{RV}{Runtime Verification}
\newacronym[longplural={Systems-on-Chip}]{soc}{SoC}{System-on-Chip}
\newacronym{srga}{SRGA}{self-reconfigurable gate array}
\newacronym{tessla}{TeSSLa}{Temporal Stream-based Specification Language}
\newacronym{gpu}{GPU}{Graphics Processing Unit}

\usepackage[backend=biber,maxbibnames=99]{biblatex}
\newcommand{\field}[1]{\textit{#1}}
\newcommand{\func}[1]{\textbf{#1}}
\AtBeginEnvironment{quote}{\itshape}
\newcommand{\engine}{Dryas\xspace}

\begin{document}

\title{Dryas: A Reprogrammable Engine for High-Speed Interconnect Tracing and Analysis}


\author{Manuel Bröchin}
\authornote{Work was done at ETH Zürich}
\affiliation{
	\institution{SBB}
	\country{Switzerland}
}
\email{maneul.broechin@sbb.ch}

\author{Tom Kuchler}
\orcid{0009-0002-8091-0313}
\affiliation{
	\institution{ETH Zürich}
	\country{Switzerland}
}
\email{tom.kuchler@inf.ethz.ch}

\author{Michael Giardino}
\authornotemark[1]
\orcid{0000-0002-9906-720X}
\affiliation{
	\institution{Huawei}
	\country{Switzerland}
}
\email{michael.giardino@huawei.com}

\author{David Cock}
\authornotemark[1]
\orcid{0000-0003-2997-6560}
\affiliation{
	\institution{Neutrality}
	\country{Switzerland}
}
\email{david@neutrality.ch}

\author{Timothy Roscoe}
\orcid{0000-0002-8298-1126}
\affiliation{
	\institution{ETH Zürich}
	\country{Switzerland}
}
\email{trocoe@inf.ethz.ch}

\begin{abstract}
The proliferation of heterogeneous components in modern computing systems has 
been accompanied by new higher bandwidth and lower latency interconnects. 
These interfaces and protocols are enormously complex and the process of developing, 
debugging, and analyzing FPGA-based implementations requires significant engineering work.
Moreover, once a functional implementation is completed, optimization of the controller 
and associated software requires processing potentially hundreds of gigabytes of trace data.

In this paper, we present \engine, an open source tool for analyzing such an interconnect. 
We developed our tool, using minimal hardware resources, alongside an FPGA implementation of a very high speed, low latency 
(30~GiB/s, 200~ns) interconnect. 
With our run-time reprogrammable overlay engine we can inspect this interconnect to find 
rare, complex, or transient events even at full operation. 
This filtering engine is based on non-deterministic finite automata (NFAs), efficiently 
implemented using state transition elements (STEs), allowing us to trace events at a cache-line
granularity. 
Moreover we can change the filters in less than a second, without reprogramming the FPGA or
interfering with the running application. 
This data enables not only debugging the implementation of the interconnect itself, but 
analyzing the behavior of accelerated applications.

We examine the mathematical basis for using NFAs and describe their implementation 
on a real coherent CPU-FPGA research platform. 
We then evaluate the scalability of \engine for various size NFAs, followed by two different 
use cases: debugging FPGA implementation of the interconnect and analyzing cache behavior.

\end{abstract}

\maketitle


\glsresetall

\section{Introduction} \label{sec:introduction}

Modern computer architecture has been moving towards more specialized and heterogeneous hardware topologies.
While smart network interface cards \cite{Mellanox:2020} and GPUs are the most prevalent examples, \glspl{fpga} have also seen significant use as accelerators.
Their increased flexibility in execution unit design and host system interaction
leads to a number of interesting topologies.
Aside from traditional PCIe-attached \gls{fpga} accelerators such as 
Xilinx Alveo \cite{alveou250,alveou280} which tend to mirror a GPU-inherited
batch-processing model, there is increasing interest in more tightly-coupled CPU+FPGA
systems \cite{crockett_zynq_2014,ccix,cxl,Choi:2019:IDAM}.
This tight coupling generally comes with the advantages of high bandwidth, low latency,
and the possibility of fine-grained memory management including caching and coherence.
However, implementation of not only application software but the protocols themselves
require performant systems to analyze low-level interactions.

High-bandwidth interconnects produce enormous amounts of data in very short times,
making finding a specific message or sequence difficult.
While capturing all data and then performing offline analysis is possible, just the boot process
can create traces of hundreds of gigabyes, necessitating fast storage connected via a
channel with equivalent bandwidth.

For low-level tracing data to be useful, there must be a mechanism to filter only relevant messages.
Moreover, a filter should be easily reconfigurable, allowing for different filters to
be quickly loaded, even into a running system.
If reconfiguration takes too long, or worse, requires resynthesis of the FPGA design, it
severely limits the usefulness of it for active debugging and analysis.

We present \engine, a run-time reprogrammable filter engine that transparently taps into
the FPGA implementation of the interconnect interface in order to extract specific
sequences of messages at a cache-line granularity. 
In order to meet the requirements of line-rate high-bandwidth throughput, complex event
filtering, and fast reconfigurability, we look to existing \gls{fpga} stream processing
tools for inspiration. 
\glspl{nfa} composed of \glspl{ste} allow for the creation of complex, expressive
filtering queries that can be reconfigured in less than a second without reprogramming
the FPGA or even stopping application execution.
\engine uses a minimal amount of FPGA resources and on our test system can filter data at 
a 30~GiB/s line rate.
\engine can not only give insight into existing communication
protocols and assist in the extension and improvement of them,  
it can also be deployed alongside running applications on the \gls{fpga}, 
to perform realtime, fine-grained analysis of the application's use of the protocol.

\gls{fpga} vendors provide some hardware tracing tools in their software platforms that support low-level debugging of interconnects.
For instance, Xilinx ILAs allow observing arbitrary signals on the \gls{fpga} and
specifying trigger conditions as state machines of these signal values.
However, they have significant shortcomings that make them unsuitable for our needs.
First, ILAs are connected to the remote machine over a low-bandwidth JTAG interface that
is not suitable to stream out large amounts of data.
Second, the state machines are limited to $16$ states with one accepting state, and each
state has fan-out at most $3$, severely limiting the trigger condition complexity.

In \autoref{sec:background}, we provide background information on the hardware
platform under test, \glspl{nfa} and overlays for \glspl{fpga}.
We then examine the design and implementation in Sections \ref{sec:design} and
\ref{sec:implementation}.
\autoref{sec:evaluation} evaluates the \engine's scalability and examines
two use cases: an exploration of cross-socket latency at the protocol layer
and cache behavior of a simple FPGA benchmark.

\section{Background} \label{sec:background}

The design of \engine is motivated by the need to analyse and debug the
\gls{eci}~\cite{Ramdas:CCKit:2025}.  In \autoref{sec:background:problem} we
briefly introduce the relevant features of the interconnect, and the
challenges found in such a system.  We elaborate the crucial design choices
that result from these features, then examine \glspl{nfa} and overlay
architectures and how they can be used for analysis.

\subsection{Problem Statement} \label{sec:background:problem}

Enzian \cite{enzian}~\cite{Cock:2022:Enzian} is a heterogeneous research server with a 48-core
Marvell \gls{thx} CPU connected to a Xilinx Virtex Ultrascale+ \gls{fpga} via
\gls{eci}, a 30~GiB/s coherent interconnect.
\gls{eci} connects the \gls{thx}'s L2 cache to the \gls{fpga}.

\gls{eci} comprises 2 links, each consisting of 12 x 10~Gb/s serial lines, using a 3-layer protocol.  
At the lowest layer, an Interlaken-like protocol~\cite{interlaken} provides reliable delivery of a bidirectional datastream between the endpoints.

 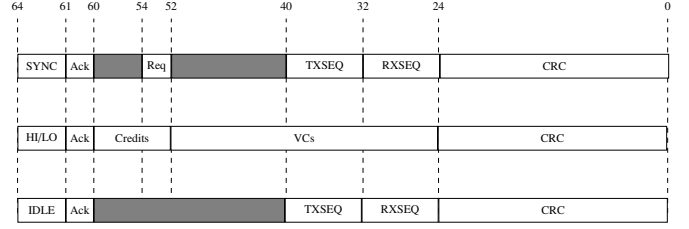
\begin{figure}[t!]
 	\centering
 	\scriptsize
 	\resizebox{0.5\textwidth}{!}{
 		\begin{tikzpicture}
 			\tikzstyle {field} = [draw, rectangle, minimum height=0.5cm, anchor=west, fill=white];
 			\def\bitwidth{0.2cm}
			
 			\newcommand\bitline[2][0]{
				
 				\begin{scope}[on background layer]
 					\draw (#1 |- idle.south) edge [dashed, shorten <= 0.2cm] (#1 |- lim);
 					\node [label=above:#2] at (#1 |- lim) {};
 				\end{scope}
 			}
			
 			\node [field, minimum width=5*\bitwidth] (sync) at (0,0) {SYNC};
 			\node [field, minimum width=1*\bitwidth, right=0 of sync.east] (ack) {Ack};
 			\node [field, minimum width=5*\bitwidth, right=0 of ack.east, fill=gray] (empty_1) {};
 			\node [field, minimum width=3*\bitwidth, right=0 of empty_1.east] (req) {Req};
 			\node [field, minimum width=12*\bitwidth, right=0 of req.east, fill=gray] (empty_2) {};
 			\node [field, minimum width=8*\bitwidth, right=0 of empty_2.east] (txseq) {TXSEQ};
 			\node [field, minimum width=8*\bitwidth, right=0 of txseq.east] (rxseq) {RXSEQ};
 			\node [field, minimum width=24*\bitwidth, right=0 of rxseq.east] (crc) {CRC};
			
 			\node [field, below=1cm of sync, minimum width=5*\bitwidth] (hi_lo) {HI/LO};
 			\node [field, minimum width=1*\bitwidth, right=0 of hi_lo.east] (data_ack) {Ack};
 			\node [field, minimum width=8*\bitwidth, right=0 of data_ack.east] (credits) {Credits};
 			\node [field, minimum width=28*\bitwidth, right=0 of credits.east] (vcs) {VCs};
 			\node [field, minimum width=24*\bitwidth, right=0 of vcs.east] (crc) {CRC};

 			\node [field, below=1cm of hi_lo, minimum width=5*\bitwidth] (idle) {IDLE};
 			\node [field, minimum width=1*\bitwidth, right=0 of idle.east] (idle_ack) {Ack};
 			\node [field, minimum width=20*\bitwidth, right=0 of idle_ack.east, fill=gray] (empty_3) {};
 			\node [field, minimum width=8*\bitwidth, right=0 of empty_3.east] (idle_txseq) {TXSEQ};
 			\node [field, minimum width=8*\bitwidth, right=0 of idle_txseq.east] (idle_rxseq) {RXSEQ};
 			\node [field, minimum width=24*\bitwidth, right=0 of idle_rxseq.east] (crc) {CRC};
			
 			\coordinate (lim) at (0,1);
			
 			\bitline[sync.west]{64};
 			\bitline[sync.east]{61};
 			\bitline[ack.east]{60};
 			\bitline[req.west]{54};
 			\bitline[req.east]{52};
 			\bitline[txseq.west]{40};
 			\bitline[rxseq.west]{32};
 			\bitline[crc.west]{24};
 			\bitline[crc.east]{0};
			
 		\end{tikzpicture}
 	}
 	\caption{The four types of block layer headers: sync, hi/lo data, and idle blocks. The first three bits indicate the type of the block: $\text{SYNC} = 110$, $\text{HI} = 100$, $\text{LO} = 101$ and $\text{IDLE} = 111$.}
 	\label{fig:block:headers}
 	
 \end{figure}

The middle (block) layer divides this reliable stream into fixed-size 64~B blocks, each of which carries an 8~B header and seven 8~B data words. 

There are four types of blocks: sync, high-data, low-data, and idle blocks, indentified by the first $3$~b of its header.
The structure of each type of block header is depicted in \autoref{fig:block:headers}.
Sync blocks carry only meta information in the header and are used to synchronize the two nodes after an error condition or at startup.
High- and low-data blocks carry higher-layer messages in the seven data words.
Idle blocks are sent when neither data nor sync blocks are sent in order to keep the link active and synchronized.
Every header contains a 24-bit CRC checksum field to provide error detection.
Moreover, both endpoints acknowledge the reception of data blocks using the dedicated \field{Ack} field in all headers.

Using block-level metadata, this layer provides link monitoring and initialization, credit-based flow control, and the multiplexed transmission of independent
\glspl{vc}. 

The third, \gls{vc} layer is responsible for interpreting the messages for each of the 14 \glspl{vc} (See \autoref{tab:vcs}) and performing the corresponding action.
Each of the 14 \glspl{vc} carries a different message type (e.g. forward request, or read response), and it is the responsibility of the top layer to correctly interpret the per-\gls{vc} packet format~\cite{jakob} and route and process individual messages correctly.
The \field{VCs} field of data blocks indicates the recipient \gls{vc} for each of the seven data words. 
The block layer consumes \gls{vc}-layer messages from the \gls{vc} layer and puts them into data blocks.
Because \gls{vc}-layer messages have variable size, messages may be split across multiple blocks.
Thus, the block layer is also responsible for re-assembly of \gls{vc} messages at the receive side before handing them up to the \gls{vc} layer.
Messages on different \glspl{vc} are used differently.
Their header information and payload size thus varies depending on the type of message~\cite{jakob}.

\begin{table}
  \centering
  \rowcolors{2}{white}{gray!40}
  \begin{tabular}{|r|l|c|}
    \hline
    VC id & Message types & Acronym \\
    \hline
    0 & I/O request & IREQ \\
    1 & I/O response & IRSP \\
    2,3 & Memory request w data & MREQ \\
    4,5 & Memory response w data & MRSP \\
    6,7 & Memory request w/o data & MREQ \\
    8,9 & Memory forward request & MFWD \\
    10,11 & Memory response w/o data & MRSP \\
    12 & Multiplexed co-processor data & MXC \\
    13 & Multicore debugging \& link data & MDLD \\
    \hline
  \end{tabular}
  \caption{Virtual channels as provided by the block layer. }
  \label{tab:vcs}
\end{table}

A detailed account of all message types and their meaning can be found in previous work \cite{jakob, Ramdas:CCKit:2025}.
We are most interested in memory coherency messages which are handled by the \gls{dirc} module using \glspl{vc} 2 to 11.

Given the 300~MHz system clock of Enzian the number of blocks transmitted per cycle and direction is just under $1.5$, while the number of \gls{vc} layer
messages transported by these blocks is limited to $10.5$ per cycle per direction --- a little over 6 billion messages per second. 
\engine is intended to monitor messages at either the block or the \gls{vc} layer.

\engine filters the high-speed stream of messages transmitted over \gls{eci}
using an \gls{nfa} engine.  While stream processing using \glspl{fpga} is
common, our setting imposes unique constraints: We want to understand the
protocol implementation itself.  As a consequence, we are interested in the
interactions \emph{between} observed messages.  In other words, we look for
patterns in the sequence of messages.  This is an important difference
compared to common applications of stream processing on \glspl{fpga}, many of
which are either interested in pattern matching in a stream of text (e.g.
regex matching \cite{reconfigurable_regex_fpga}, network intrusion detection
\cite{complexeventdetection, bit_split}, log processing \cite{hawk, hare}) or
in detecting complex predicates within a single message (e.g. high-frequency
trading \cite{high_frequency_trading}, SQL query processing
\cite{flexible_query_processor, streams_on_wires}).

To understand the interaction between messages, we need to detect the
\emph{intent} or \emph{effect} of an individual message.  We extract this as a
predicate over the message header and send only the relevant information to
the \gls{nfa}.  Thus \engine has two stages: first evaluating predicates on
headers, then driving an \gls{nfa} with these predicates to implement the
filter.  The block layer and the different \gls{vc} protocols have different header
structures and require different predicates.  Moreover, the predicates may be
complex and contain pieces of information from multiple headers.  The input
decoding in each \gls{ste} must thus be very flexible.  On the other hand, the
\glspl{nfa} expressing our filters contain a relatively small number of
states: The \gls{nfa}'s complexity depends largely on the underlying
protocol's complexity.  Compared to regex search in text, with
arbitrarily-large patterns, the \gls{eci} protocols consist of comparatively
short sequences of interactions, thus the patterns themselves have moderate
size.

\subsection{Nondeterministic Finite Automata (NFAs)} \label{sec:background:nfa}
We use \glspl{nfa} as a means to filter streams of messages.
The goal is to take an input stream of messages $(m_1, m_2, \dots)$ and output a substream of it according to a filter rule.
Formally, the \gls{nfa} consists of a set of states $S$ and a transition function $\delta$ for state updates.
Two special subsets of the states are \emph{initial states} $S_0$, which are active at the start,
and \emph{accepting states} $F$, which indicate when the \gls{nfa} \emph{accepts} a message (indicating it should be added to the output stream).
The \emph{transition function} defines the conditions for a state to become \emph{active}.
For each message that is read from the input stream, the transition function is executed to
find new active states from current ones.
If any states in $F$ are activated by the arrival of a message, this message is added to the output stream.

\begin{figure}
  \footnotesize
  \centering
    \begin{tikzpicture}
      \tikzstyle {state} = [draw, circle, minimum size = 1cm];
      \tikzstyle {cell} = [draw=none, rectangle, minimum height = 0.5cm, minimum width = 0.4cm];
      \def\dist{2cm}

      \node [state] (1) at (0,0) {$s_0$};
      \node [state] (2) at (0,\dist) {$s_1$};
      \node [state] (3) at (\dist, \dist) {$s_2$};
      \node [state, minimum size = 0.9cm] at (3) {};

      \draw (1) edge [->, transform canvas={xshift=-0.2cm}] node [midway, label=left:a] {} (2);
      \draw (2) edge [->, transform canvas={xshift=0.2cm}] node [midway, label=right:a] {} (1);
      \draw (2) edge [->, transform canvas={yshift=-0.2cm}] node [midway, label=below:b] {} (3);
      \draw (3) edge [->, transform canvas={yshift=0.2cm}] node [midway, label=above:b] {} (2);
      \draw (1) edge [->, out = 45, in = 315, looseness=5] node [midway, label=left:b] {} (1);
      \draw (3) edge [->, out = 315, in = 225, looseness=5] node [midway, label=above:a] {} (3);
      \draw [<-] (1.south) --++ (0, -0.5cm);

      \begin{scope}[shift={(4,1.5)}]

        \foreach \x\i\s\o in {0/ /$s_0$/ ,1/a/$s_1$/ , 2/a/$s_0$/ , 3/a/$s_1$/ ,
          4/b/$s_2$/b, 5/a/$s_2$/a, 6/a/$s_2$/a, 7/b/$s_1$/ , 8/a/$s_0$/ , 9/a/$s_1$/ , 10/b/$s_2$/b }
        {
          \node [cell] (t_\x) at (\x*0.4cm, 0) {\x};
          \node [cell, below=0 of t_\x] (i_\x) {\i};
          \node [cell, below=0 of i_\x] (s_\x) {\s};
          \node [cell, below=0 of s_\x] (o_\x) {\o};
          \draw (t_\x.north west) -- (o_\x.south west);
        }

        \node [cell, anchor=east, left=.15cm of t_0] {time:};
        \node [cell, anchor=east, left=.15cm of i_0] {input:};
        \node [cell, anchor=east, left=.15cm of s_0] {state:};
        \node [cell, anchor=east, left=.15cm of o_0] {output:};
      \end{scope}
    \end{tikzpicture}
  \caption{The \gls{nfa} is graphically depicted on the left. Input, active state and filtered output are listed ordered by time on the right. }
  \label{fig:nfa:filter}
\end{figure}
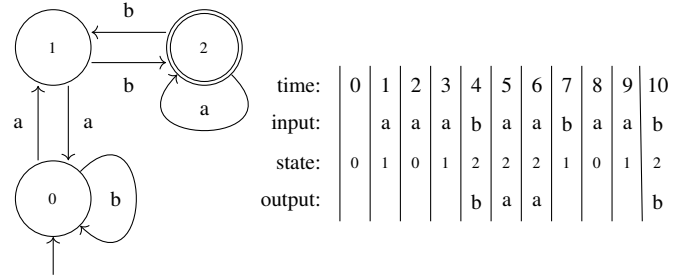

\autoref{fig:nfa:filter} shows an example \gls{nfa} corresponding to the formal description in \autoref{eq:exp:nfa}.
The right side of \autoref{fig:nfa:filter} shows how this \gls{nfa} filters a stream of abstract messages.

\begin{align}
	\label{eq:exp:nfa}
  \begin{split}
  \Sigma = \{&a, b\} \\
  S = \{ &s_0, s_1, s_2\} \\
  \delta = \{&(s_0, a) \to \{s_1\}, (s_0, b) \to \{s_0\}, (s_1, a) \to \{s_0\}, \\
             &(s_1, b) \to \{s_2\},(s_2, a) \to \{s_2\}, (s_2, b) \to \{s_1\}\} \\
  s_0 = &s_0 \\
  F = \{&s_2\}
\end{split}
\end{align}

Alphabet $\Sigma$ is the set of all inputs the \gls{nfa} recognizes.
In our use case this is a set of all possible predicates of message headers (e.g.
message types).
The predicates are extracted from the messages outside of the \gls{nfa} and allow re-purposing the \gls{nfa} to recognize use case-specific properties.

\glspl{nfa} allow composing basic conditions on predicates to more complicated patterns.
Formally, these patterns are called regular languages over the alphabet $\Sigma$.
By implementing the pattern matching using \glspl{nfa}, which can have multiple states
active at the same time, as opposed to \glspl{dfa}, where only one state is active at any
time, we can make use of the parallel hardware of the \gls{fpga}.
Since an \gls{nfa} uses less states than an equivalent \gls{dfa}, the hardware cost of using \glspl{nfa} is smaller than that of using \glspl{dfa}.


\begin{figure}[t!]
	\centering  
	\footnotesize
	\begin{subfigure}[t]{0.24\textwidth}
		\centering
		\begin{tikzpicture}
			\tikzstyle {node} = [draw, circle, minimum size = 1cm];
			\def\dist{2cm}
			
			\node [node] (s0) at (0,0) {$s_0$};
			\node [node] (s1) at (\dist, 0) {$s_1$};
			\node [node, minimum size = 0.9cm] at (s1) {};
			\draw [<-] (s0.south) --++ (0,-0.5cm);
			\draw (s0) edge [->, transform canvas={yshift=0.2cm}] node [midway, label=above:{a,b}] {} (s1);
			\draw (s1) edge [->, transform canvas={yshift=-0.2cm}] node [midway, label=below:a] {} (s0);
			\draw (s1) edge [->, out = 315, in = 225, looseness=5] node [midway, label=above:b] {} (s1);
		\end{tikzpicture}
		\caption{General \gls{nfa}}
		\label{fig:general:nfa}
	\end{subfigure}
	\hfill
	\begin{subfigure}[t]{0.24\textwidth}
		\centering
		\begin{tikzpicture}
			\tikzstyle {node} = [draw, circle, minimum size = 1cm];
			\def\dist{2cm}
			
			\node [node] (s0) at (0,0) {$s_0$};
			\node [node] (s1) at (\dist,0.7) {$s_{1,1}$};
			\node [node, minimum size = 0.9cm] at (s1) {};
			\node [node] (s2) at (\dist,-0.7) {$s_{1,2}$};
			\node [node, minimum size = 0.9cm] at (s2) {};
			\draw [<-] (s0.south) --++ (0,-0.5cm);
			\draw (s0) edge [->, bend left] node [midway, label=above:{a,b}] {} (s1);
			\draw (s1) edge [->, bend left] node [midway, label=above:{a}] {} (s0);
			\draw (s1) edge [->] node [midway, label=right:{b}] {} (s2);
			\draw (s2) edge [->] node [midway, label=below:{a}] {} (s0);
			\draw (s2) edge [->, out = 315, in = 225, looseness=5] node [midway, label=above:{b}] {} (s2);
		\end{tikzpicture}
		\caption{\gls{hnfa} equivalent to \autoref{fig:general:nfa}}
		\label{fig:homogeneous:nfa}
	\end{subfigure}
	\caption{Correspondence between general \gls{nfa} and homogeneous \gls{nfa}}
\end{figure}
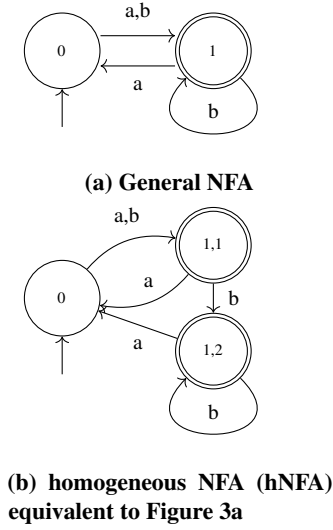

Runtime configurable \glspl{nfa} are commonly implemented on \glspl{fpga} using an overlay architecture \cite{reconfigurable_regex_fpga, napoly_2}:
The basic blocks, often called \glspl{ste} in the context of \glspl{nfa}, are overprovisioned and statically connected with each other at synthesis time.
Each \gls{ste} implements the functional behavior of a single state of the \gls{nfa}.
Logic inside an \gls{ste} allows representing different transition functions, often by matching the input symbol against some configurable memory.
To reduce the logic within a single \gls{ste}, \glspl{nfa} are commonly transformed into
homogeneous form before mapping them to the overlay:
In homogeneous form, all transitions into a state trigger on the same set of inputs.
This ensures that each \gls{ste} needs to
check only a single transition function for all its neighbors.
Importantly, any \gls{nfa} can be represented by a \gls{hnfa}, by duplicating
some states.
An example of a \gls{nfa} and the equivalent \gls{hnfa} is illustrated in \autoref{fig:general:nfa} and \autoref{fig:homogeneous:nfa} respectively.
Static connections in the network of \glspl{ste} can be switched on and off to implement \gls{nfa} with a different graph structures.
The static hardware layout requires the design to trade off hardware resources for the generality of the \glspl{nfa}.
We will go further into detail about our \gls{ste} implementation in \autoref{sec:design:stes}

\subsection{Related Work} \label{sec:background:overlays}

Karakchi et al. \cite{napoly_1,napoly_2} use an \gls{ste}-based overlay architecture to implement Napoly, a large-scale, general-purpose \gls{nfa} engine on \glspl{fpga}.
Napoly is targeted towards regex search over a one-byte symbol alphabet such as ASCII text, thus the 
input decoding is a single lookup table with 256 entries.
Moreover, to support large regex patterns, the design is optimized 
for packing as many \glspl{ste} as possible onto an \gls{fpga}.
Napoly achieves packing 24k states on a Stratix 5 GX 7A \gls{fpga}.
As a consequence of the dense and large design, routing data to and 
between \glspl{ste} becomes challenging and the design bandwidth
is limited to 90~MB/s.
Thus Napoly is unsuitable for us because we require different trade-offs:
Instead of many \glspl{ste} with a limited input decoding and low bandwidth,
we require few \glspl{ste} that can consume composed predicates from message 
headers in each cycle and compute complex transition functions on these predicates.

Teubner et al. \cite{skeleton_automata_fpga} implement XML projection on \glspl{fpga}.
Their approach has at its core a runtime configurable \gls{nfa} engine implemented as an overlay architecture.
The individual hardware components making up the overlay graph, 
called segments, allow matching arbitrary XML tags.
Because of the hierarchical structure of XML, projection specifiers are 
strictly linear, so each segment selects an XML node-set based 
on the selection of its sole predecessor segment.
Thus the overlay architecture has the very simple path structure.
Since we want to allow general \glspl{nfa}, this simple 
structure is not suitable for our work.
Moreover, the input decoding, while more complex than in Napoly, is 
restricted to string matching.
Our input decoding will need to match on message headers, or predicates 
derived from headers, which contain multiple pieces of information.
We thus need to allow more complex matching mechanisms such as 
combinations of string matching and logical operators.

\section{Design} \label{sec:design}

\engine processes incoming \gls{eci} data in three conceptual steps:
First, predicates are extracted from the \gls{eci} message header.
Second, each \gls{ste} computes its transition function on these 
predicates and update their state accordingly.
Third, based on whether the \gls{nfa} is in an accepting state, 
the input data is added to the output or discarded.

The concrete predicates that are extracted from message headers in step one 
must be specified at synthesis time and depend on the application.
We will discuss two concrete examples of extracted predicates 
in \autoref{sec:implementation}. 

We will now go into more detail on how transition functions are computed
and states are updated..
In \autoref{sec:design:overlay} we first describe the rings-of-cliques 
graph, a parametrizable graph we use as the structure of our overlay.
In \autoref{sec:design:stes} we describe how we implement reconfigurable 
\glspl{ste} as the basic blocks of our overlay architecture.

\subsection{Overlay Graph} \label{sec:design:overlay}

The overlay is a static graph of \glspl{ste} that is fixed at synthesis time.
An \gls{hnfa} is then implemented by the overlay at runtime by 
configuring each \gls{ste}.
The process of implementing a \gls{hnfa} on the overlay is called \emph{mapping}, whereby 
each state of the \gls{hnfa} is mapped to exactly one \gls{ste}.
Therefore, we need to ensure that adjacent states are mapped to adjacent \glspl{ste}.
The synthesized overlay graph limits the expressibility of the \glspl{nfa} (i.e. filters)
in two ways:
First, the number of \glspl{ste} determines the maximum number of \gls{hnfa} states.
Second, \gls{ste} connectivity limits \gls{hnfa} connectivity.
It is therefore crucial to allow the user to flexibly trade-off number of states and density of the overlay graph at synthesis time.

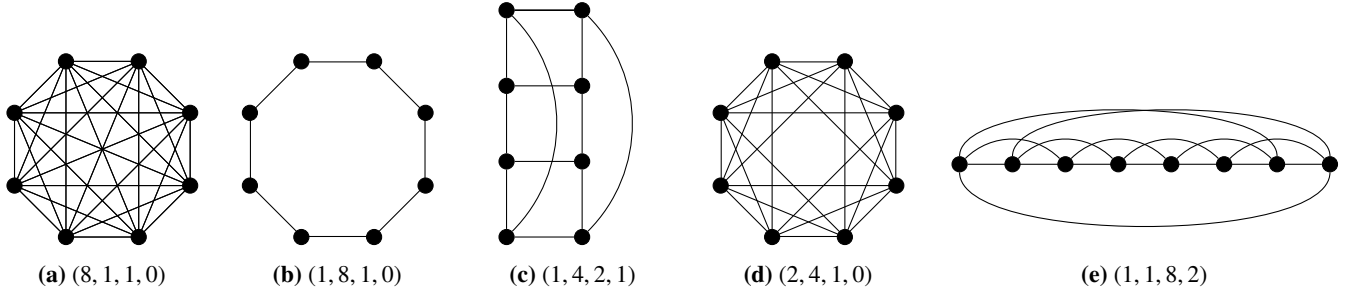
\begin{figure*}
  \centering
  \footnotesize

  \begin{subfigure}[t]{0.15\textwidth}
    \centering
    \begin{tikzpicture}
      \tikzstyle{state} = [draw, circle, minimum size = 0.2cm, fill=black, inner sep=0]

      \node[draw=none,minimum size=2.5cm,regular polygon,regular polygon sides=8] (poly) {};

      \foreach \x in {1,...,8}
      {
        \node [state] at (poly.corner \x) {};
        \foreach \y in {1,...,8}
        {
          \draw (poly.corner \x) -- (poly.corner \y);
        }
      }
    \end{tikzpicture}
    \caption{$(8, 1, 1, 0)$}
    \label{fig:config:graphs:1}
  \end{subfigure}
  \hfill
  \begin{subfigure}[t]{0.15\textwidth}
    \centering
    \begin{tikzpicture}
      \tikzstyle{state} = [draw, circle, minimum size = 0.2cm, fill=black, inner sep=0]

      \node[draw,minimum size=2.5cm,regular polygon,regular polygon sides=8] (poly) {};

      \foreach \x in {1,...,8}
      {
        \node [state] at (poly.corner \x) {};
      }
    \end{tikzpicture}
    \caption{$(1, 8, 1, 0)$}
    \label{fig:config:graphs:2}
  \end{subfigure}
  \hfill
  \begin{subfigure}[t]{0.15\textwidth}
    \centering
    \begin{tikzpicture}
      \tikzstyle{state} = [draw, circle, minimum size = 0.2cm, fill=black, inner sep=0]
      \def\n{4}
      \def\m{2}

      \pgfmathtruncatemacro{\lx}{ \n - 1 }%
      \pgfmathtruncatemacro{\ly}{ \m - 1 }%

      \foreach \x in {1,...,\n} {
        \foreach \y in {1,...,\m} {
          \node [state] (s_\x_\y) at (\y, \x) {};
        }
      }
      \foreach \x in {1,...,\lx} {
        \pgfmathtruncatemacro{\a}{ \x + 1 }%
        \draw (s_\x_\m) -- (s_\a_\m);
        \foreach \y in {1,...,\ly} {
          \pgfmathtruncatemacro{\b}{ \y + 1 }%
          \draw (s_\x_\y) -- (s_\a_\y);
          \draw (s_\x_\y) -- (s_\x_\b);
          \draw (s_\n_\y) -- (s_\n_\b);
        }
      }
       \foreach \y in {1,...,\m}{
         \draw (s_1_\y) edge [out=45, in=315] (s_\n_\y);

      }
    \end{tikzpicture}
    \caption{$(1, 4, 2, 1)$}
    \label{fig:config:graphs:3}
  \end{subfigure}
  \hfill
  \begin{subfigure}[t]{0.15\textwidth}
    \centering
    \begin{tikzpicture}
      \tikzstyle{state} = [draw, circle, minimum size = 0.2cm, fill=black, inner sep=0]

      \node[draw,minimum size=2.5cm,regular polygon,regular polygon sides=8] (poly) {};

      \foreach \x in {1,...,8} {
        \node [state] at (poly.corner \x) {};
        \foreach \y in {1,...,2}{
          \pgfmathtruncatemacro{\z}{ Mod(\x + \y, 8) + 1  }%
          \draw (poly.corner \x) -- (poly.corner \z);
        }
      }
    \end{tikzpicture}
    \caption{$(2, 4, 1, 0)$}
    \label{fig:config:graphs:4}
  \end{subfigure}
  \hfill
  \begin{subfigure}[t]{0.3\textwidth}
    \centering
    \begin{tikzpicture}
      \tikzstyle{state} = [draw, circle, minimum size = 0.2cm, fill=black, inner sep=0]

      \foreach \x in {0,...,7} {
        \pgfmathsetmacro{\xpos}{ \x * 0.7 }%
        \node [state] (\x) at (\xpos, 0) {};
      }
      \draw (0) -- (7);
      \draw (0) edge [out=270, in=270, looseness=0.5] (7);

      \foreach \x in {0,...,5} {
        \pgfmathtruncatemacro{\b}{ Mod(\x + 2, 8) }%
        \draw (\x) edge [out=45, in=135] (\b);
      }
      \draw (0) edge [out=90, in=90, looseness=0.5] (6);
      \draw (1) edge [out=90, in=90, looseness=0.5] (7);

    \end{tikzpicture}
    \caption{$(1, 1, 8, 2)$}
    \label{fig:config:graphs:5}
  \end{subfigure}
  \caption{Various configurations (C,L,R,N) of the rings-of-cliques overlay graph.}
  \label{fig:ring:of:cliques}
\end{figure*}

The overlay graph chosen to allow this trade-off is a \emph{rings-of-cliques (RoC)} graph.
An instance of an RoC graph is defined by $4$ parameters:
clique size \emph{C}, ring length \emph{L}, number of rings \emph{R} 
and number of neighboring rings \emph{N}.
$(C,L,R,N)$ denotes a particular configuration.
Each vertex $v_{c,l,r}$ in the structure is uniquely identified by 
indices $c \in [C],l \in [L],r \in [R]$ denoting the node's index in the clique, 
the clique's index in the ring and the ring's index.
Two vertices $v_{c,l,r}$ and $v_{c',l',r'}$ are connected if either of the 
following is true
\begin{itemize}
\item Their rings are not more than $N$ apart and their cliques are in the 
same position of their respective ring. This includes the case where two 
vertices are in the same clique. More formally, if $l = l'$ and 
$\min(r-r'\bmod R, r'-r \bmod R) \leq N$
\item They are in neighboring cliques on the same ring or formally, if $r = r'$ 
and $\min(l-l'\bmod L, l'-l\bmod L) = 1$
\end{itemize}
\autoref{fig:ring:of:cliques} illustrates several configurations of the RoC overlay graph.

The problem of mapping each state to an \gls{ste}, such that the neighborship 
relation is preserved is commonly referred to as subgraph mapping problem.
The subgraph mapping problem is well-known to be NP-complete.
We formulate the subgraph mapping problem as an \gls{ip} and solve it using standard
\gls{ip} solvers such as Gurobi \cite{gurobi} or Coin-OR \cite{coinor}.
Other authors use heuristics to speed up the mapping from \gls{nfa} to the overlay graph
\cite{napoly_1,micron_ap}.
As mentioned earlier, we expect the \gls{hnfa} size of most filters to be 
small to moderate (at most several $100$ states), therefore mapping can easily be performed as a standard \gls{ip} formulation.

\subsection{State Transition Elements} \label{sec:design:stes}

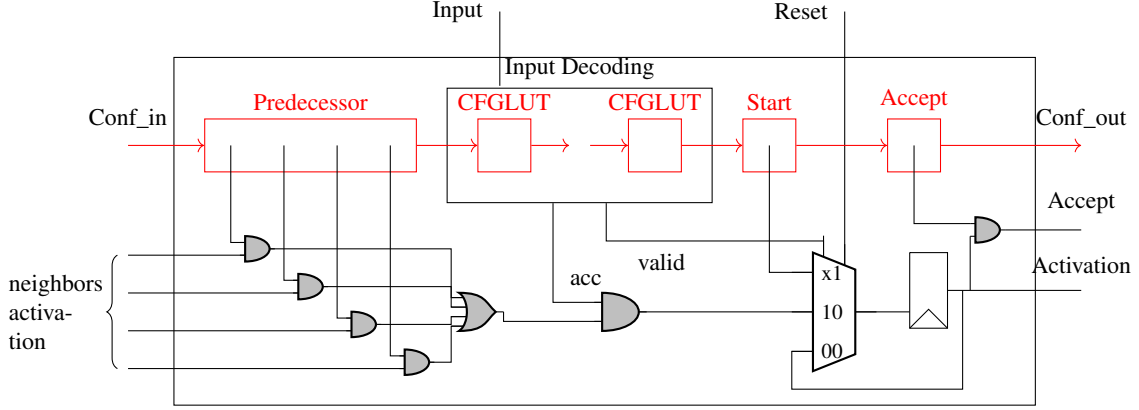
\begin{figure*}
  \centering
  \small
    \begin{tikzpicture}
      \def\dist{0.8cm}
      \def\cfgdist{0.5cm}
      \def\regsize{0.5cm}

      \tikzstyle{config} = [draw, rectangle, red, inner sep = 0cm, minimum size = 0.7cm];
      \tikzstyle{node} = [draw, rectangle, inner sep = 0.4cm, minimum height = 0.7cm];
      \tikzstyle{cfgchain} = [draw, ->, red];
      \tikzstyle{signal} = [draw];
      \ctikzset{
        logic ports/scale=0.3,
        logic ports/fill=lightgray
      }
      \tikzset{demux/.style={muxdemux, muxdemux def={Lh=2.8, Rh=2, w=1, NL=3, NB=0, NT=2, NR=1}}}

      \node [config, label=above:\color{red} Predecessor, minimum width = 2.8cm] (pred) at (0,0){};
      \node [config, right = \dist of pred, label=above:\color{red} CFGLUT] (cfg1) {};
      \node [red, right = \cfgdist  of cfg1] (dots) {$\dots$};
      \node [config, right = \cfgdist of dots, label=above:\color{red} CFGLUT] (cfg2) {};
      \node [config, right = \dist of cfg2, label=above:\color{red} Start] (start) {};
      \node [config, right = 1.2cm of start, label=above:\color{red} Accept] (accept) {};

      \node [node, fit={(cfg1) (dots) (cfg2)}, label=above:Input Decoding] (input_decoding) {};

      \node [or port, scale=1.5, number inputs = 4, below = 1.95cm of cfg1.west] (or) {}; 
      \node [and port, scale=1.5, right = 1.2cm of or] (and) {};  
      \node [demux, right=2.2cm of and] (mux) {};
      \node [right] at (mux.blpin 1) {\footnotesize x1};
      \node [right] at (mux.blpin 2) {\footnotesize 10};
      \node [right] at (mux.blpin 3) {\footnotesize 00};

      \coordinate [right=1cm of mux.north] (reg_1) {};
      \coordinate [right=0.25cm of reg_1] (reg_0) {};
      \coordinate [right=0.5cm of reg_1] (reg_2) {};
      \coordinate [below=1cm of reg_2] (reg_3) {};
      \coordinate [below=0.5cm of reg_2] (reg_6) {};
      \coordinate [below=1cm of reg_1] (reg_4) {};
      \coordinate [below=0.75cm of reg_0] (reg_5) {};
      \path [draw] (reg_1) -- (reg_2) -- (reg_3) -- (reg_4) -- (reg_1);
      \path [draw] (reg_3) -- (reg_5) -- (reg_4);

      \coordinate [right=0.2cm of reg_6] (split) {};
      \coordinate [above=0.8cm of split] (join) {};
      \node [and port, right = 0.1cm of join] (and2) {};

      \foreach \x\y\z in {0.25/1.2cm/1, 0.5/1.7cm/2, 0.75/2.2cm/3, 1/2.7cm/4} {
        \node[and port, below = \y of $(pred.west) !\x! (pred.east)$] (and_\z) {};
        \draw ($(pred.west) !\x-0.125! (pred.east)$) |- (and_\z.bin 1);
        \draw [signal] (and_\z.bout) -| (or.in \z) -- (or.bin \z);
      }

      \node [node, fit={(and2) (and_4) (input_decoding) (pred)}] (ste) {};

      \coordinate [left = 0.6cm of ste] (left_coord) {};
      \coordinate [below = 0.6cm  of ste] (below_coord) {};
      \coordinate [right = 0.6cm  of ste] (right_coord) {};
      \coordinate [above = 0.6cm  of ste] (up_coord) {};

      \coordinate [below =0.5cm of mux.lpin 3] (mux_cord) {};

      \node [inner sep=0] (inp) at ($(input_decoding.north west) !0.2! (input_decoding.north east)$) {};
      \draw (inp) -- (inp |- up_coord);
      \draw [signal] ($(input_decoding.south west) !0.4! (input_decoding.south east)$) |- (and.bin 1) node [midway, label=above right:acc] {};
      \path [draw] ($(input_decoding.south west) !0.6! (input_decoding.south east)$) --++ (0,-0.5cm) node [midway, inner sep=0.3cm, label=below right:valid] {} -| (mux.btpin 1);
      \draw [signal] (or.out) |- (and.bin 2);
      \draw [signal] (and.bout) |- (mux.blpin 2);
      \draw (start.center) |- (mux.blpin 1);
      \draw [signal] (mux.brpin 1) -- (reg_1 |- mux.rpin 1);
      \draw [signal] (reg_6) -- (right_coord |- reg_6) {};
      \path [signal] (split) |- (mux_cord) |- (mux.blpin 3);
      \draw (accept.center) |- (and2.in 1);
      \draw (and2.out) -- (right_coord |- and2.out);
      \draw (mux.btpin 2) -- (up_coord -| mux.tpin 2);

      \draw (split -| and2.in 2) |- (and2.in 2);

      \node [label=left:Input] at (inp |- up_coord) {};
      \node [label=left:Reset] at (mux.btpin 2 |- up_coord) {};
      \node [label=above:Conf\_in] at (pred.west -| left_coord) {};
      \node [label=above:Conf\_out] at (accept.east -| right_coord) {};
      \node [label=above:Accept] at (and2.out -| right_coord) {};
      \node [label=above:Activation] at (reg_6 -| right_coord) {};
      \draw [decorate,decoration={brace,amplitude=4pt,raise=4pt}]
      (and_4.in 2 -| left_coord) -- (and_1.in 2 -| left_coord) node [midway, xshift=-10pt, label={[text width=1cm] left:{neighbors activation}}] {};

      \draw [cfgchain] (left_coord |- pred.west) -- (pred.west);
      \draw [cfgchain] (pred) -- (cfg1);
      \draw [cfgchain] (cfg1) -- (dots.west);
      \draw [cfgchain] (dots.east) -- (cfg2);
      \draw [cfgchain] (cfg2) -- (start);
      \draw [cfgchain] (start) -- (accept);
      \draw [cfgchain] (accept.east) -- (right_coord |- accept.east);

      \foreach \x in {1,2,3,4} {
        \draw [signal] (left_coord |- and_\x.in 2) -- (and_\x.bin 2);
      }

    \end{tikzpicture}
  \caption{Logic diagram of a single \gls{ste} with $4$ predecessor \glspl{ste}.}
  \label{fig:runtime:reconfiguration}
\end{figure*}

Given a mapping of \gls{nfa} states to \glspl{ste}, we need to configure each 
\gls{ste} such that it implements the behavior of its state.
As discussed, each \gls{ste} represents one state of an \gls{hnfa}.
The functional behavior of a state in an \gls{hnfa} is the following
\begin{quote}
  A state activates if either it is an initial state and time $t = 0$, or one of its predecessors was active at $t-1$ and it accepts the input of time $t$.
  A state de-activates at time $t$ if it doesn't activate at time $t$ and it receives valid input.
\end{quote}
The restriction on de-activating is unique to our design and was added because the
interfaces considered often have valid traffic only on a subset of all clock cycles.

In order to allow runtime configuration of \engine, the following four components 
of the overlay need to be configurable at runtime:
(1) The input decoding of the \gls{ste} that computes the transition function 
on the input predicate, (2) the \field{start} register value that determines if 
an \gls{ste} is a starting state, (3) whether or not a state is accepting, and 
(4) the routing between \glspl{ste}.

This behavior is implemented to be runtime reconfigurable as depicted in \autoref{fig:runtime:reconfiguration}. 
Red are registers used for runtime reconfigurations, black are data path elements.
We shift a bitstring into the configuration registers at runtime to control the 
functional behavior of the \gls{ste}.
At time $t$, we choose one out of three possible values as the next activation:
If reset is asserted, we choose the \field{start} register value.
Else, if no valid input was received in this cycle, as indicated by the 
\field{valid} output of \texttt{Input Decoding}, we keep the previous activation.
Else, we check if any transition into this state is active.
A transition is active at time $t$ if (1) the input of time $t$ is accepted by 
\texttt{Input Decoding}, i.e. if \field{acc} is asserted, and (2) if a predecessor 
state was active at time $t-1$.
We output the activation of this state at time $t$ to all connected \glspl{ste}, 
which use it to compute their activation at time $t+1$.
Further, if this \gls{ste} corresponds to an accepting state of the \gls{nfa}, 
as indicated by the \field{Accept} configuration register, we assert the \field{Accept} 
output that is aggregated centrally by the \gls{nfa} engine.

The \gls{ste} design so far is relatively standard.
The main difference to the related work is in the \texttt{Input Decoding} module.
As discussed before, the main challenge of our work comes from wanting to interpret 
complex predicates extracted from one or multiple headers.
Moreover, we want to be able to consider multiple protocols, where the structure 
and contents of the headers change with the protocol.

The concrete predicates depend on the message headers of the interface protocol 
under consideration and the predicate extraction performed on these headers.
As discussed above, this predicate extraction is specified at synthesis time.
Similarly, the structure of the configurable input decoding logic of the \gls{ste} 
needs to be fixed at synthesis time as it will generally depend on the shape and 
content of the extracted predicates.
However, the basic structure of the \texttt{Input Decoding} is always composed of 
the \gls{cfglut} Xilinx primitive:
A \gls{cfglut} implements an arbitrary 5-to-1 logic function.
It can be configured at runtime by sequentially pushing a $32$-bit configuration string.
The loading of a config string is synchronized by a clock, but the input-output behavior of the \gls{cfglut} is combinatorial.
The output of the 5-bit input $x \in \{0,\dots,31\}$ is given by bit $x$ of the configuration.
The \texttt{Input Decoding} uses \glspl{cfglut} to match on parts of the message headers.
Which parts of the message headers are matched depends on the protocol and use case.
Moreover, we often want to aggregate the outputs of different \glspl{cfglut} in order to represent more complex transition functions.
We will discuss two examples in \ref{sec:implementation}.

Now, instead of one input stream, \engine should be able to process $n$
interleaved input streams at the same time with the same filter.
This is useful for independently filtering coherency messages addressing different 
\glspl{cl} (see \autoref{sec:evaluation:cachemiss}). 
Conceptually, this can be achieved by replicating the overlay architecture $n$ times,
mapping the same \gls{nfa} onto each overlay, and de-multiplexing the input stream into
$n$ input streams, one per overlay.
Interleaved stream processing can also be accomplished with less hardware duplication 
by generalizing the \glspl{ste} in a way that allows them to keep track of $n$ 
activations concurrently.
Instead of receiving one activation per neighboring \gls{ste}, now $n$ activations are received.
Similarly, the input decoding computes an $n$-vector where entry $i$ indicates that the
predicate matches on substream $i$.
The implementation of the \gls{ste} displayed in \ref{fig:runtime:reconfiguration} stays
mostly the same, with the difference that some of the signals are now $n$-bit buses and
some of the gates additionally need to aggregate $n$-bit buses to signals or broadcast
signals to $n$-bit buses.


\section{Implementation} \label{sec:implementation}
%
%
%

We will now examine how we apply the design discussed in \autoref{sec:design} to concrete use cases.
To reduce the width of the configurable input decoding logic, we perform 
data reduction on the incoming data by extracting only the relevant predicates of the message headers.
This preprocessing is fixed at compile time and thus determines a specific 
set of queries that can be expressed at runtime using the input decoding logic.
This specialization of \engine reduces flexibility of expressible patterns,  but 
the reduced data width improves scalability of the design in two ways: 
(1) the input decoding logic can be made slimmer, and (2) fewer routing 
resources are needed.
Moreover, it allows us to enrich the data sent to the \gls{nfa} with derived 
predicates, described below.
Now we describe the data reduction performed on block and \gls{vc} layers and the resulting
query languages.
Note that for simplicity, we will describe the structure and behavior of a 
single link, but for full \gls{eci} these structures are duplicated.

\subsubsection{Block Layer}

 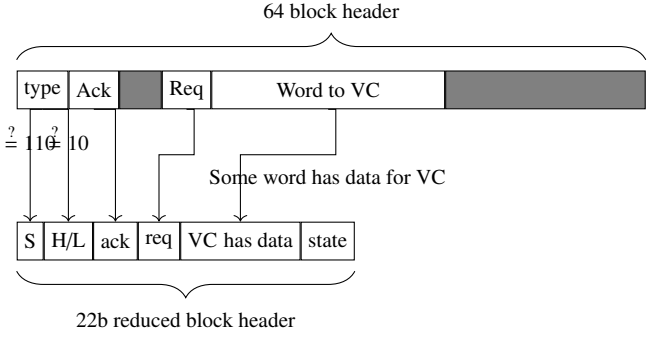
\begin{figure}[t!]
   \centering
   \footnotesize
     \begin{tikzpicture}
       \tikzstyle {field} = [draw, rectangle, minimum height=0.5cm, anchor=west, fill=white];
       \def\bitwidth{0.11cm}

       \node [field, minimum width=5*\bitwidth] (type) at (0,0) {type};
       \node [field, minimum width=1*\bitwidth, right=0 of type.east] (ack) {Ack};
       \node [field, minimum width=5*\bitwidth, right=0 of ack.east, fill=gray] (empty_1) {};
       \node [field, minimum width=3*\bitwidth, right=0 of empty_1.east] (req) {Req};
       \node [field, minimum width=28*\bitwidth, right=0 of req.east] (vcs) {Word to \gls{vc}};
       \node [field, minimum width=24*\bitwidth, right=0 of vcs.east, fill=gray] (empty_2) {};

       \node [field, minimum width = 1*\bitwidth] (issync) at (0,-2cm) {S};
       \node [field, minimum width = 1*\bitwidth, right=0 of issync] (isdata) {H/L};
       \node [field, minimum width = 1*\bitwidth, right=0 of isdata] (ack2) {ack};
       \node [field, minimum width = 2*\bitwidth, right=0 of ack2] (req2) {req};
       \node [field, minimum width = 14*\bitwidth, right=0 of req2] (vc) {$\text{\gls{vc}}_i$ has data};
       \node [field, minimum width = 3*\bitwidth, right=0 of vc] (state) {state};

       \draw [decorate,decoration={brace,amplitude=10pt,mirror,raise=4pt}]
       (issync.south west) -- (state.south east) node [midway, yshift=-14pt, label=below:{$22$b reduced block header}] {};
       \draw [decorate,decoration={brace,amplitude=10pt,raise=4pt}]
       (type.north west) -- (empty_2.north east) node [midway, yshift=14pt, label=above:{$64b$ block header}] {};

       \node [below=0.5cm of vcs.south, label=below:{Some word has data for $\text{\gls{vc}}_i$}] (compvc) {};
       \node [below=0.5cm of req.south] (compreq) {};

       \draw [->] (type.south) -| node [midway, label=below:{$\overset{?}{=} 110$}] {} (issync);
       \draw [->] (type.south) -| node [midway, label=below:{$\overset{?}{=} 10x$}] {} (isdata);
       \draw [->] (ack.south) -| (ack2);

       \path[draw, ->] (vcs.south) -| (compvc.east) -| (vc);
       \path[draw, ->] (req.south) -| (compreq.east) -| (req2);

     \end{tikzpicture}
   \caption{Data reduction performed at the block layer with both incoming and outgoing block header.}
   \label{fig:block:datareduction}
 \end{figure}

The block layer receives an incoming and outgoing block each cycle, leaving \engine with two 64~b headers per cycle.
This data is further reduced to a 44~b input containing important fields of the
block headers, specifically, the header type, the \gls{vc} 
indices of data blocks, and control bits.
Additionally, the input contains 
the estimated states of the input and output links which are derived from the observed
headers within the data reduction module.

The structure and content of the reduced input is depicted in \autoref{fig:block:datareduction}.

The input decoding logic then allows the formulation of an arbitrary logic function
on 8~b from the input: 4~b each from the incoming and outgoing block headers.
The selection of 2 x 4~b is implemented in one logic slice (see \cite{ultrascale_clb}).
The runtime configurable 8-to-1 logic function is implemented by multiplexing between
8~\glspl{cfglut}, also in a single logic slice.
We will see examples of these transition functions in \ref{sec:evaluation:latency}.

\subsubsection{VC Layer}

The 10 cache coherency \glspl{vc} allow exchanging necessary messages for
implementing a directory-based MOESI-style cache coherence protocol.
Messages can be grouped into memory requests, forward requests,
responses, and eviction notifications containing different information in their 
headers.
However the most interesting fields are common to all coherency headers:
\begin{itemize}
	\item The \emph{opcode} of the \gls{eci} message (bits $[63,59]$).
	\item The memory \emph{address} affected by the \gls{eci} message (bits $[39,7]$).
\end{itemize}
In contrast to the block layer, where an incoming and an outgoing block are processed 
in every clock cycle, messages on each \gls{vc} are received whenever they have been 
fully assembled on the block layer.
The interface between block and \gls{vc} layer consists of message \glspl{fifo}, one 
\gls{fifo} per \gls{vc} per direction.
\engine directly taps this interface and therefore must be able to handle a full set
of 10 incoming and 10 outgoing \gls{vc} messages in a cycle.
The input to \engine at the \gls{vc} layer consists of a \emph{message batch} each cycle
\begin{equation} \label{eq:msg:batch}
  \left(m_i^{\text{in}}\right)_{i=0}^{9}\left(m_i^{\text{out}}\right)_{i=0}^{9}
\end{equation}
where $m_i^{\text{in/out}}$ is a valid \gls{eci} message for coherency \gls{vc} $i$, or all ones if there is no data on the \gls{vc} this cycle.
From each message in the batch we extract opcode and address field and pass them to the \gls{nfa}.
By sending these fields in the order of the messages in the batch, we implicitly also pass the \emph{\gls{vc}} and the \emph{sender} information to the \gls{nfa}.

\begin{figure}[t!]
  \centering
  \footnotesize
    \begin{tikzpicture}
      \def\bitlen{0.13cm}
      \tikzstyle{bitfield} = [draw, rectangle, anchor=north west, minimum height = 0.5cm]

      \node [bitfield, minimum width = 5*\bitlen] (opcode) at (0,0) {op};
      \node [bitfield, minimum width = 19*\bitlen, right=0 of opcode, fill=gray] (e1) {};
      \node [bitfield, minimum width = 23*\bitlen, right=0 of e1] (addr_hi) {addr[32:$n+1$]};
      \node [bitfield, minimum width = 10*\bitlen, right=0 of addr_hi] (addr_lo) {addr[$n$:0]};
      \node [bitfield, minimum width = 7*\bitlen, right=0 of addr_lo, fill=gray] (e2) {};
      \draw [decorate,decoration={brace,amplitude=10pt,raise=4pt}]
      (opcode.north west) -- (e2.north east) node [midway, yshift=14pt, label=above:{$64$b \gls{eci} header}] {};

      \node [below=1cm of e1.south east, label={above:{$\text{mode1} \lor \text{addr}[32:n+1] = A$}}] (comp) {};

      \node [bitfield, minimum width = 5*\bitlen] (opcode2) at (0,-2cm) {op};
      \node [bitfield, minimum width = 1*\bitlen, right=0 of opcode2] (v) {v};
      \node [bitfield, minimum width = 10*\bitlen, right=0 of v] (addr_lo2) {idx};
      \draw [decorate,decoration={brace,amplitude=10pt,mirror,raise=4pt}]
      (opcode2.south west) -- (addr_lo2.south east) node [midway, yshift=-14pt, label=below:{$(n+7)$b reduced \gls{eci} header}] {};

      \draw [->] (opcode) -- (opcode2);
      \draw [->] (addr_lo) |- node [midway, label=below:{$\text{mode2} \land \text{addr}[n:0]$}] {} (addr_lo2.east);
      \path [draw,->] (addr_hi) |- (comp.east) -| (v);

    \end{tikzpicture}
  \caption{Predicate extraction performed at the \gls{vc} layer for mode1, where \gls{cl} addresses are ignored and mode2, where $2^{n+1}$ separate substreams are observed, distinguished by \gls{cl} address.}
  \label{fig:datareduction}
\end{figure}
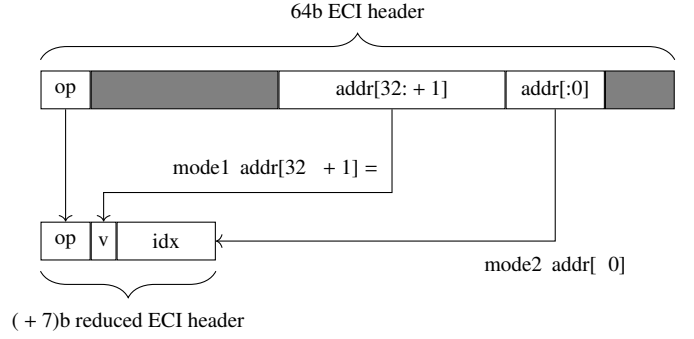

The input decoding implemented in the \glspl{ste} allows expressing the following transition 
functions on the extracted predicates of a message batche $M$:
\begin{enumerate}
\item A basic transition function checks opcode, sender and \gls{vc} of all messages in the
message batch $M$ and accepts if there is a message $m \in M$ that matches the selection.
\item \func{Any}($\text{P}^1, \dots, \text{P}^k$): is true if any of the basic transition functions $\text{P}^1, \dots, \text{P}^k$ match on $M$.
\item \func{None}($\text{P}^1, \dots, \text{P}^k$): is true if none of the basic transition functions $\text{P}^1, \dots, \text{P}^k$ matches on $M$.
\end{enumerate}
This is implemented using one \gls{cfglut} for each of the ten headers sent by both \gls{fpga} and CPU.
Each \gls{cfglut} can match the opcode against an arbitrary pattern.
By correctly configuring the individual \glspl{cfglut} and aggregating their partial matches using either logical Or or And, we can implement the full set of transition functions above.
We will see examples in \ref{sec:evaluation:cachemiss}.

Since caches maintain separate state for each \gls{cl}, we are often interested in
tracking and filtering the messages for each \gls{cl} individually.
However, some use cases such as simple message filters, may ignore
\gls{cl} information entirely.
\engine thus supports the following two modes of dealing with \gls{cl} address field:
(1) ignoring \gls{cl} information or
(2) partitioning the \gls{eci} message stream into $n$ disjoint substreams, based on \gls{cl} address and filtering each substream independently

The first mode feeds all input messages to the same \gls{nfa}.
The second mode allows the user to specify a base \gls{cl} address and a number $n$ of
subsequent \glspl{cl} that should be tracked.
The stream of messages is then partitioned into $n$ substreams based on the $\log n$
lowest bits of the \gls{cl} index.
Conceptually, each of the $n$ substreams in the user-specified range is then fed to
a different instance of the same \gls{nfa} (see: \autoref{sec:design:stes}).

\section{Evaluation} \label{sec:evaluation}

%
%

We first evaluate how \engine scales with overlay size and filter
complexity in \autoref{sec:evaluation:scalability}, and the show
\engine's versatility in two use-cases: cross-socket memory latency
(\autoref{sec:evaluation:latency}) and cache-miss characterization
(\autoref{sec:evaluation:cachemiss}). 

These two examples demonstrate the filtering engine's utility for understanding the underlying system.
Moreover, they show that \engine is versatile enough to allow inspecting properties at both the block and the \gls{vc} layer.

\subsection{Scalability} \label{sec:evaluation:scalability}

\begin{figure*}[t!]
	\includegraphics[width=\textwidth]{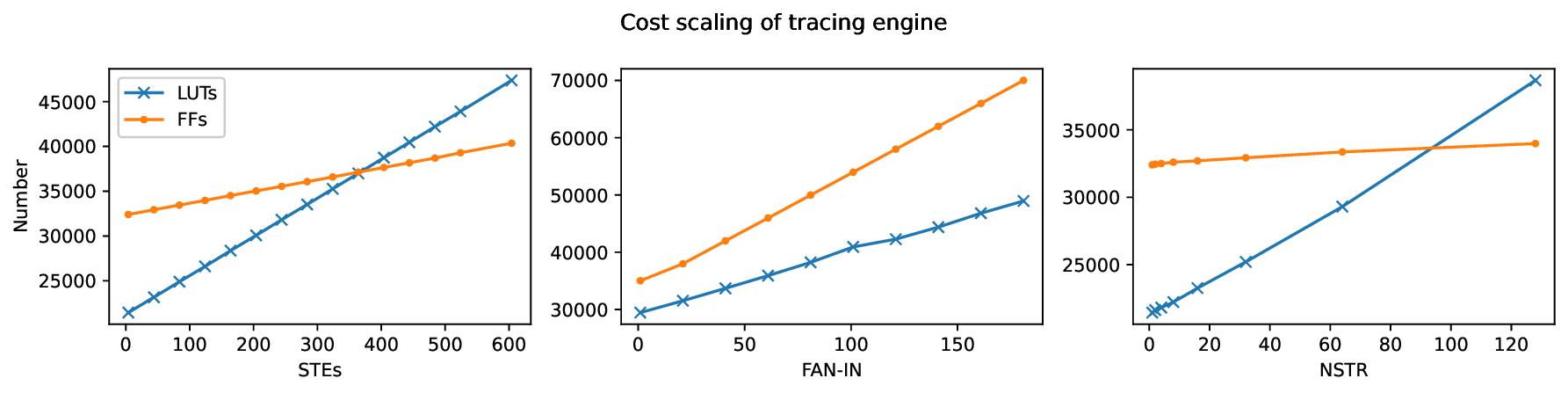}
	\caption{Hardware cost of \engine measured in \glspl{lut} and \glspl{ff} when increasing number of \glspl{ste} (STEs), fan-in per state (FAN-IN) and the number of filtered substreams (NSTR).}
	\label{fig:plot:scaling}
\end{figure*}

We show the hardware cost in \glspl{lut} and \glspl{ff} when
synthesizing \engine for a given overlay.  We use a Xilinx Virtex
Ultrascale+ with 1.1~M \glspl{lut} and 2.3~M \glspl{ff}.  \engine has
a fixed base cost of \engine accounts for about 20k \glspl{lut} and
29k \glspl{ff}, most of which (18k \glspl{lut} and 23k \glspl{ff}) due
to the AXI XDMA IP core used over PCIe to configure \engine and output
the trace; using \gls{eci} and FPGA DRAM would reduce this
significantly.

\autoref{fig:plot:scaling} plots \engine's hardware cost as measured in \glspl{lut} and \glspl{ff}.
The base configuration in the left and right plots has $4$ states, fan-in $4$ and considers $1$ unified stream.
The base configuration in the middle plot has 200 \glspl{ste}, such that the connectivity can be increased without modifying the number of \glspl{ste}.
To examine the scaling from these base configurations we scale along three dimensions: number of \glspl{ste}, the fan-in of each \gls{ste}, and the number of filtered substreams (NCLS).
In the left plot, we increase the number of \glspl{ste} without increasing their connectivity (increasing $L$).
In the middle plot, we increase the connectivity between \glspl{ste} (increasing $N$).
In the right plot we increase the copies of the filtering \gls{nfa} that are executed on de-multiplexed substreams.
In all cases, the cost scales approximately linearly with the increase of the overlay parameter, allowing the user to trade-off these parameters at will.
The cost of \engine scales well and has a small cost, taking up less than 2.17\% of the
\glspl{lut} and less than 3.64\% of \glspl{ff} for a configuration with 600 states, even 
when using the XDMA IP.

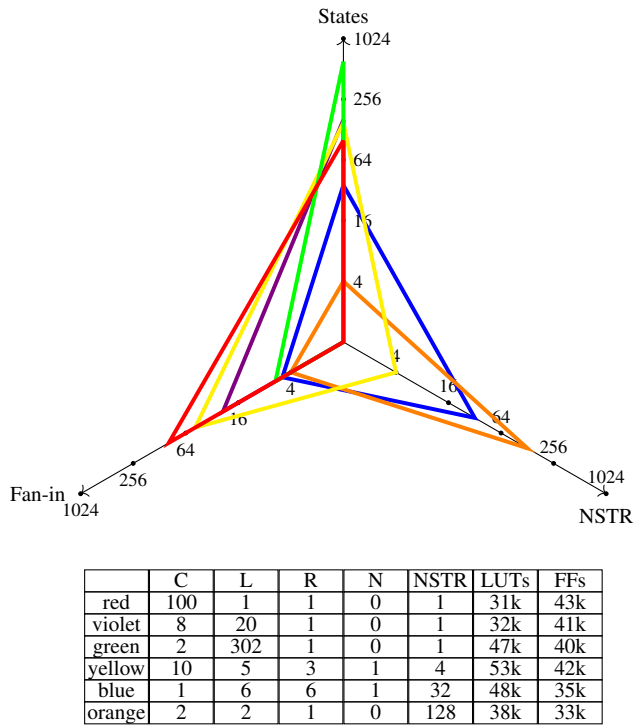
\begin{figure}[t!]
  \centering
  \footnotesize
  \begin{tikzpicture}

    \def\maxstates{10}
    \def\maxdeg{10}
    \def\maxncls{10}

    \def\raylen{8cm}
    
    \node[draw=none,minimum size=\raylen,regular polygon,regular polygon sides=3] (poly) {};
    \foreach \x\lab\dir in {1/States/above, 2/Fan-in/left, 3/NSTR/below}
    {
      \draw (poly.center) edge [->] (poly.corner \x);
      \node [label=\dir:\lab] at (poly.corner \x) {};
    }
    
    \foreach \x\max\dir in {1/\maxstates/right, 2/\maxdeg/below, 3/\maxncls/above}
    {
      \foreach \l in {1,...,5}
      {
        \pgfmathsetmacro{\f}{\l / 5}%
        \pgfmathtruncatemacro{\val}{pow(2, \l * \max/5)}%
        \node [draw, circle, inner sep = 0, minimum size = 0.05cm, fill=black, label=\dir:{\scriptsize{\textcolor{black}{\val}}}] at ($(poly.center) !\f! (poly.corner \x)$) {};
      }
    }

    \newcommand\spider[4][]{
      \pgfmathsetmacro{\fr}{log2(#2) / \maxstates}%
      \coordinate (cr) at ($(poly.center) !\fr! (poly.corner 1)$) {};
      \pgfmathsetmacro{\fl}{log2(#3) / \maxdeg}%
      \coordinate (cl) at ($(poly.center) !\fl! (poly.corner 2)$) {};
      \pgfmathsetmacro{\fc}{log2(#4) / \maxncls}%
      \coordinate (cc) at ($(poly.center) !\fc! (poly.corner 3)$) {};

      \path [draw, line width=1.5pt, #1] (cr) -- (cl) -- (cc) -- (cr);
    }
    \spider[blue]{36}{5}{32};
    \spider[orange]{4}{4}{128};
    \spider[violet]{160}{24}{1};
    \spider[yellow]{150}{50}{4};
    \spider[green]{604}{6}{1};
    \spider[red]{100}{100}{1};

    \begin{scope}[shift={(0,-3)}]
      \matrix (mA) [matrix of nodes,row sep=-\pgflinewidth, inner sep=0, text depth=0.5ex,text height=1.8ex,nodes in empty cells, nodes={rectangle,draw, minimum width=3em},anchor=north]
      {
         & C & L & R & N & NSTR & LUTs & FFs \\
        red & 100 & 1 & 1 & 0 & 1 &  31k & 43k \\
        violet & 8 & 20 & 1 & 0 & 1 & 32k & 41k \\
        green & 2 & 302 & 1 & 0 & 1 & 47k & 40k \\
        yellow & 10 & 5 & 3 & 1 & 4 & 53k & 42k \\
        blue & 1 & 6 & 6 & 1 & 32 & 48k & 35k \\
        orange & 2 & 2 & 1 & 0 & 128 & 38k & 33k \\
      };

    \end{scope}
    
  \end{tikzpicture}

  \caption{Spider diagram showing maximal configurations that can be implemented witout timing violations.}
  \label{fig:spider}
\end{figure}

\autoref{fig:spider} shows maximal configurations we could synthesize
without timing issues.  The largest clique we can handle (red) has
$100$ \glspl{ste}, all pairwise connected, and so any \gls{hnfa} up to
$100$ states can be mapped to this configuration without application
knowledge.  
Violet and green configurations show how we can trade off connectivity
for states by decreasing clique size and increasing ring length. 
The yellow, blue and orange configurations introduce more parallel
streams: yellow and blue are plausible use cases, whereas orange shows
maximally feasible parallelism.

\autoref{fig:spider} also shows the resource usage for these
configurations, which is small in compared with the total resources
available.  \engine can thus be synthesized alongside major
applications with require most of the hardware resources of the
\gls{fpga}, while still supporting a wide variety of filtering
use-cases for them. 

\subsection{Understanding Cross-Socket Latency} \label{sec:evaluation:latency}

Next we show \engine filtered traces for relevant information,
significantly reducing the volume of output data that must be
post processed.  Our example is the \gls{eci} protocol stack,
instrumented running a memory scanning benchmark we built to 
better understand the limits of cross-socket bandwidth and latency on
Enzian.  This entails tracing actual memory transactions on the
30~GiB/s \gls{eci} link.  
In the benchmark, 16~cores on the CPU each scan
\gls{cl}-by-\gls{cl} through a different 4~kB region of \gls{fpga}
DRAM.  Each core $i$ thus scans the region [0x$i$000, 0x$(i+1)$000]. 
After each scan, the cores all synchronize with a barrier and then repeat.

\engine works here as a simple filter extracting CPU memory requests
of type \texttt{MREQ\_RLDD} (``read \gls{cl} as shared or exclusive'')
and the FPGA responses of type \texttt{MRSP\_PEMD} (``memory response
with data'').  
At the VC layer, \engine can directly match on the message opcode
field and discard messages without the desired type and sender. 
At the block layer, the type cannot directly be accessed without deep
packet inspection, but we know that \texttt{MREQ\_RLDD} arrives on
\gls{vc} 6 or 7, and \texttt{MRSP\_PEMD} is sent on \gls{vc} 4 or 5.
This information is sufficient to dramatically reduce the trace size,
enough to handle remaining false positives in a postprocessing step.
The \gls{nfa} used requires only three \glspl{ste} with fan-in 2 and
need not distinguish multiple substreams. 
A minimal overlay fitting this simple \gls{nfa} uses only 21k
\glspl{lut} and 32k \glspl{ff}, including functionality to timestamp
each message.

The VC layer tracing (not shown) shows that memory request latency at
this layer averages $0.135\mu$s, with average latency at the block
layer of $5.2\mu\text{s}$.  These are the time between arrival
of an \texttt{MREQ\_RLDD} request at the \gls{vc}/block layer and the
arrival of the corresponding \texttt{MRSP\_PEMD} response at the 
same layer, respectively.  The 40x difference in latency
shows that a large portion of available bandwidth is lost between
block and \gls{vc} layer in the \gls{eci} stack. 

\begin{figure}[t!]
  \centering
  \includegraphics[width=0.5\textwidth]{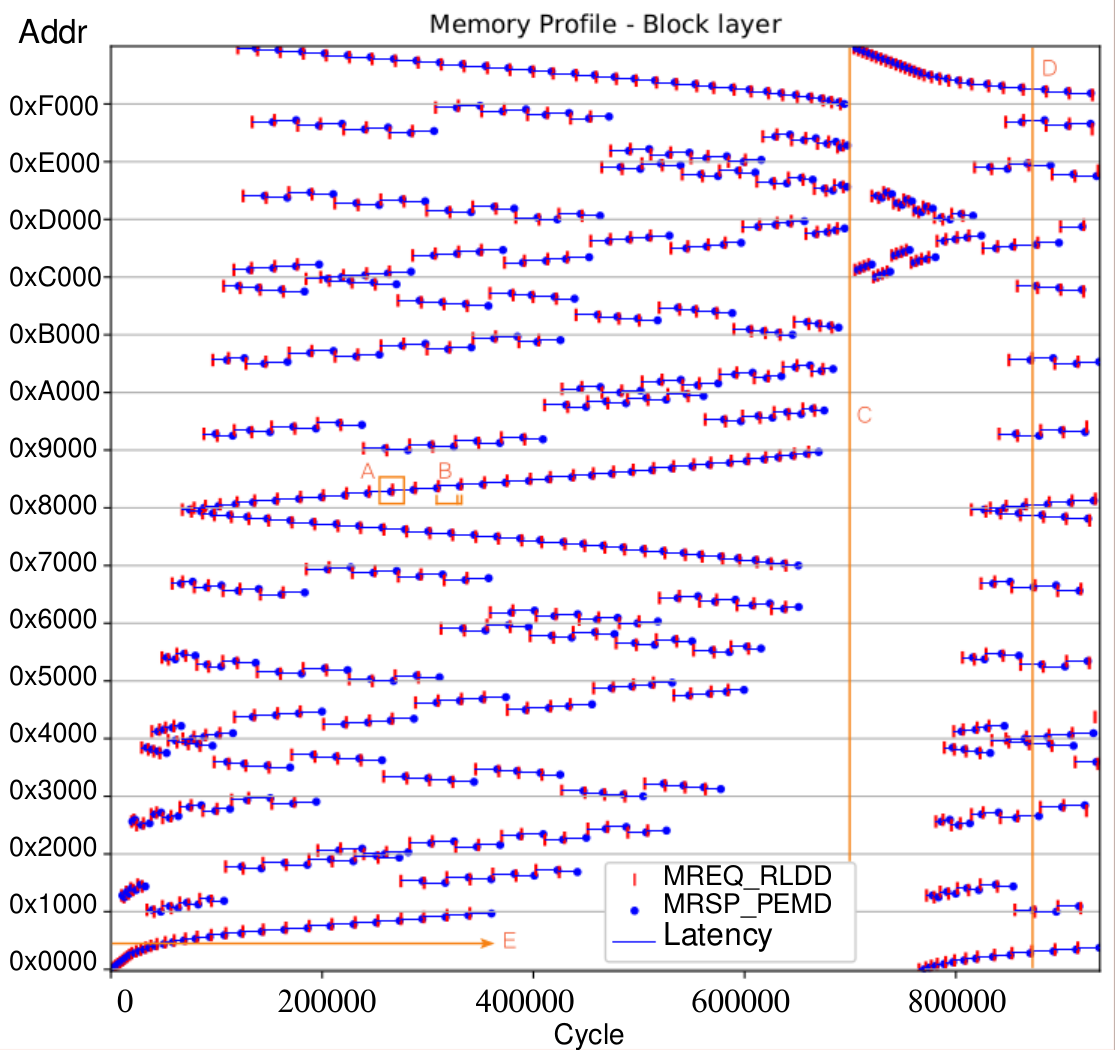}
  \caption{Memory transactions at the \gls{fpga} block layer
    during a scanning benchmark. Requests and responses are plotted
    against referenced memory address and clock cycle at which they
    were captured. Annotations are in orange.} 
  \label{fig:memory:profile:block}
\end{figure}

\autoref{fig:memory:profile:block} shows individual
\texttt{MREQ\_RLDD} and \texttt{MRSP\_PEMD} messages with timestamp on
the x-axis and the \gls{cl} address on the y-axis.  Request-response
pairs are connected with a blue horizontal line, indicating the
latency of each request at the block layer of the \gls{fpga}.

We see the single-threaded nature of the cores: each core has at most
one request outstanding, and the red \texttt{MREQ\_RLDD} markers never
cross a horizontal blue line, as in orange box ``A''. 
Overall, there are often 16 requests outstanding at the \gls{fpga},
one per core. 

Actual link latency and processing overhead on the CPU is minimal; 
cores spend most of their time waiting for the FPGA's VC layer. 
Moreover, this latency is largely due to congestion: the more
requests outstanding at the \gls{fpga}, the longer the response
time.  Observed latency is low when few cores are running and high
when many are running (seen on core 0 at ``E''), showing most of the 16 in-flight requests at the block layer are serialized at
the FPGA's \gls{vc} layer.

We also see the loss in performance incurred by the barrier synchronization between the
cores (the vertical line ``C''); some cores spend time waiting
for their peers to complete -- for example, core $0$ is waiting about
30\% of the time.  Moreover, cores don't all start immediately after
synchronization and some see a further delay, likely
due to thread scheduling. 

This analysis provides insights into the \gls{fpga} interface
implementation the system in general, and is only possible because
\engine annotates the output stream with timestamps and performs
significant on-board data reduction on the trace.  At the block layer,
we receive about 1.5 blocks of 512 bits each cycle.  The filtered
trace used for \autoref{fig:memory:profile:block} spans about $1.6
\cdot 10^9$ cycles of 300~MHz clock of the \gls{fpga}, about 5 seconds
of the benchmark.  Since blocks arrive at a constant rate, an unfiltered
trace would consist of $2.4 \cdot 10^9$ blocks 
with a total size of 1.2~TiB.  \engine reduces the size of the
captured trace by a factor of over $10^5$, largely reducing the false
positive rate, and correspondingly reducing the post processing time
required by the same factor.



\subsection{Characterizing cache misses} \label{sec:evaluation:cachemiss}


We now show the use of more complex filters and the ability to track
multiple substreams, in this case to characterize cache misses of a
given workload using the coherency messages.  

In this experiment, we store a perfectly balanced binary tree in
\gls{fpga}-attached memory and the CPU repeatedly accesses random
entries of it.
We use \engine to characterize the cache misses into the traditional
taxonomy of \emph{compulsory ($C_0$)}, \emph{capacity ($C_1$)},
\emph{conflict ($C_2$)} or
\emph{coherency ($C_3$)} misses.

 
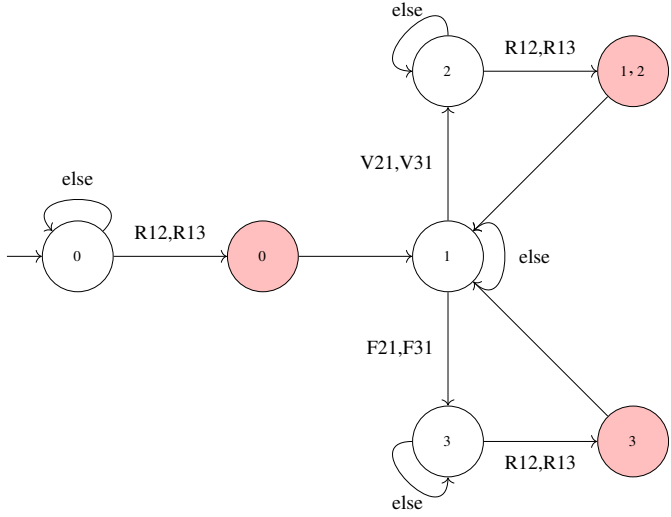
\begin{figure}
  \centering
  \footnotesize
  \resizebox{0.5\textwidth}{!}{
    \begin{tikzpicture}
      \tikzstyle{node} = [draw, circle, minimum size = 1cm]
      \def\dist{1.6cm}

      \node [node] (init) at (0,0) {$s_0$};
      \node [node, right=\dist of init, fill=red!25] (cold) {$C_0$};
      \node [node, right=\dist of cold] (eps1) {$s_1$};

      \node [node, above=\dist of eps1] (evict) {$s_2$};
      \node [node, below=\dist of eps1] (force) {$s_3$};
      \node [node, right=\dist of evict, fill=red!25] (conflict) {$C_1,C_2$};
      \node [node, right=\dist of force, fill=red!25] (coherence) {$C_3$};

      \draw[<-] (init) --++ (-1, 0);
      \draw (init) edge [->] node [midway, label=above:{R12,R13}]{} (cold);
      \draw (cold) edge [->] node [midway, label=above:{$\varepsilon$}]{} (eps1);
      \draw (eps1) edge [->] node [midway, label=left:{V21,V31}]{} (evict);
      \draw (eps1) edge [->] node [midway, label=left:{F21,F31}]{} (force);

      \draw (force) edge [->] node [midway, label=below:{R12,R13}]{} (coherence);
      \draw (evict) edge [->] node [midway, label=above:{R12,R13}]{} (conflict);
      \draw (coherence) edge [->] node [midway, label=below:{$\varepsilon$}]{} (eps1);
      \draw (conflict) edge [->] node [midway, label=above:{$\varepsilon$}]{} (eps1);


      \draw (eps1) edge [->, out=315, in=45, looseness=3] node [midway, label=right:{else}] {} (eps1);
      \draw (init) edge [->, out=45, in=135, looseness=3] node [midway, label=above:{else}] {} (init);
      \draw (evict) edge [->, out=90, in=180, looseness=3] node [midway, label=above:{else}] {} (evict);
      \draw (force) edge [->, out=180, in=270, looseness=3] node [midway, label=below:{else}] {} (force);

    \end{tikzpicture}
  }
  \caption{Life-cycle of a \gls{cl} with different miss types. States corresponding to cache misses $C_0$, $(C_1,C_2)$ and $C_3$ are red. Transition labels use an abbreviated notation for \gls{vc} layer messages: Rxy are memory requests, Fxy are forward requests, and Vxy are voluntary downgrades.}
  \label{fig:cacheline:life}
\end{figure}

\autoref{fig:cacheline:life} shows an \gls{nfa} which distinguishes
compulsory from capacity/conflict and from coherency misses on some
fixed \gls{cl}; note that we cannot distinguish capacity from conflict
misses using \gls{eci} messages alone, since we have no
information on node-local \glspl{cl} resident in a node's cache.

The first request for a \gls{cl} $x$ is a compulsory miss.
Any subsequent request to the \gls{fpga} is a witness for a further
miss, which can be categorized into capacity/conflict or coherency
based on the reason for the eviction. 
A prior voluntary eviction message (V31, V21) from the CPU
implies that the present miss is a capacity/conflict miss.
Similarly, a prior downgrade request from the \gls{fpga} (F21, F31) implies the present miss is a coherency miss.
The number of times state $C_i$ is active thus corresponds to the number of misses of type $i$.
The \gls{nfa} has $7$ states and the highest degree node $s_1$ has
fan-in 6.  Here we want to observe multiple substreams, one for each
\gls{cl}, and we synthesize an overlay with $8$ \glspl{ste} and fan-in
$6$ that can trace $32$ independent substreams.  Implementing this
uses 31k \glspl{lut} and 37k \glspl{ff}. 



We characterize the workload's cache misses by running the benchmark
three times, tracing the \gls{eci} messages with this \gls{nfa},
changing (at runtime) the set $S$ of accepting states of the 
\gls{nfa} each time: first, filtering  compulsory misses with $S =
\{C_0\}$; second, filtering capacity/conflict misses with $S =
\{(C_1,C_2)\}$; and third, filtering coherence misses with $S = \{C_3\}$.
Each resulting trace consisting of memory requests 
corresponding to the given type of cache miss.  In order to track the
state of every \gls{cl} independently, we need to track $24$
substreams, since the binary search tree used spans $24$ \glspl{cl}. 

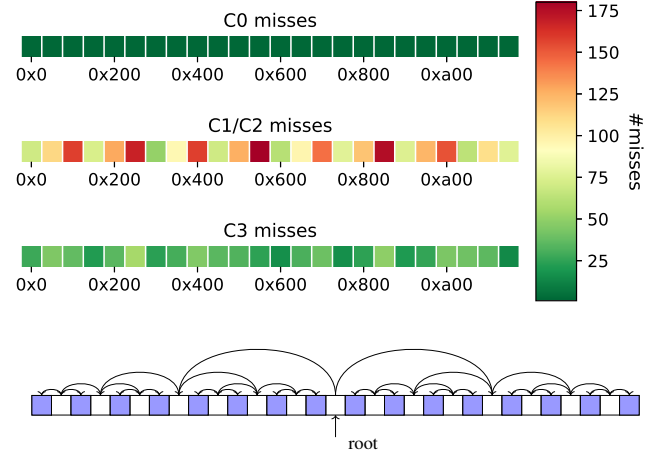
\begin{figure}
  \centering
  \Large
  \resizebox{0.5\textwidth}{!}{
    \begin{tikzpicture}
      \node [anchor=south west] (label) at (-1,2) {
        \includegraphics{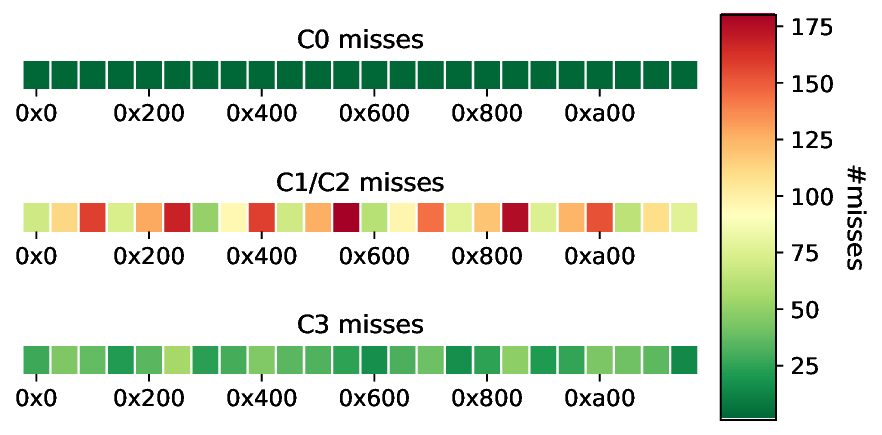}
      };


      \def\size{0.45}
      \tikzstyle {cl} = [draw, rectangle, minimum width=\size cm, minimum height=\size cm, inner sep=0];

      \foreach \i in {0,...,30} {
        \pgfmathsetmacro{\x}{ \i * \size }%
        \node [cl] (cl\i) at (\x, 0) {};
      }
      \foreach \i in {0,2,4,6,8,10,12,14,16,18,20,22,24,26,28,30} {
        \pgfmathsetmacro{\x}{ \i * \size }%
        \node [cl, fill=blue!40] (clfiller\i) at (\x, 0) {};
      }

      \foreach \f\t in {15/7, 15/23,
        7/3, 7/11, 23/19, 23/27,
        3/1, 3/5, 11/9, 11/13, 19/17, 19/21, 27/25, 27/29,
        1/0,1/2,5/4,5/6,9/8,9/10,13/12,13/14,17/16,17/18,21/20,21/22,25/24,25/26,29/28,29/30} {
        \draw (cl\f) edge [->, out=90, in=90, looseness=1] (cl\t);
      }

      \draw [draw, <-] (cl15.south) --++ (0,-0.5cm);
      \node [below=0.5cm of cl15, label=right:{root}] {};
    \end{tikzpicture}
  }
  \caption{Top: heatplot of the number of cache misses for each \gls{cl} per type. Bottom: the layout of the binary search tree in memory, explaining the miss pattern.}
  \label{fig:miss:histogram}
\end{figure}

\autoref{fig:miss:histogram} shows a heatmap of the number of cache
misses of each type per \gls{cl}.  We observe the miss pattern of each
cache miss type, and that there are many more capacity/conflict misses
than coherence misses. 

%

While this memory access behavior might be explained analytically, for
real workloads this is impossible.  In contrast, \engine's ability
to reconfigure at runtime and simultaneously trace multiple streams of
messages can help the user to understand the behavior of its workload.
Moreover, the scalability of \engine ensures that it can be used for
complex workloads. 

\section{Conclusion} \label{sec:conclusion}

We have presented \engine, a reconfigurable, low resource usage filter engine
that is easily deployed alongside other applications on an \gls{fpga}.
\engine employs a flexible overlay architecture to support
easily-reconfigurable \gls{nfa} filters.  Our design is scalable to various
sizes of \glspl{hnfa}, while maintaining low resource usage.

We have demonstrated \engine by employing it to analyze \gls{eci} in two
different scenarios: First we showed how even a simple filter can greatly
decrease analytical efforts in post processing of traces to analyze the
communication protocol.  Second we used more advanced filtering to observe
memory access patterns in a specific application and find hot spots for access
bottlenecks.

\engine is a flexible, scalable, efficient analysis and debugging tool
for live applications on modern, high-performance interconnects for
cache-coherent FPGA systems.



\printbibliography

@article{Ramdas:CCKit:2025,
	author = {Ramdas, Abishek and Cock, David and Giardino, Michael and Korolija, Dario and Ruzhanskaia, Anastasiia and Schwyn, Daniel and Turowski, Adam and Alonso, Gustavo and Roscoe, Timothy},
	title = {CCKit: An open-source toolkit for cache coherent accelerators},
	year = {2025},
	issue_date = {August 2025},
	publisher = {Association for Computing Machinery},
	address = {New York, NY, USA},
	volume = {43},
	number = {3},
	issn = {0734-2071},
	url = {https://doi.org/10.1145/3763790},
	doi = {10.1145/3763790},
	journal = {ACM Trans. Comput. Syst.},
	month = sep,
	articleno = {9},
	numpages = {30}
}

@article{complexeventdetection,
  title={Complex event detection at wire speed with FPGAs},
  author={Woods, Louis and Teubner, Jens and Alonso, Gustavo},
  journal={Proceedings of the VLDB Endowment},
  volume={3},
  number={1-2},
  pages={660--669},
  year={2010},
  publisher={VLDB Endowment}
}

@inproceedings{reconfigurable_regex_fpga,
  title={Runtime parameterizable regular expression operators for databases},
  author={Istv{\'a}n, Zsolt and Sidler, David and Alonso, Gustavo},
  booktitle={2016 IEEE 24th Annual International Symposium on Field-Programmable Custom Computing Machines (FCCM)},
  pages={204--211},
  year=2016,
  organization={IEEE}
}

@article{flexible_query_processor,
  title={Flexible query processor on FPGAs},
  author={Najafi, Mohammadreza and Sadoghi, Mohammad and Jacobsen, Hans-Arno},
  journal={Proceedings of the VLDB Endowment},
  volume={6},
  number={12},
  pages={1310--1313},
  year={2013},
  publisher={VLDB Endowment}
}

@inproceedings{skeleton_automata_fpga,
  title={Skeleton automata for FPGAs: reconfiguring without reconstructing},
  author={Teubner, Jens and Woods, Louis and Nie, Chongling},
  booktitle={Proceedings of the 2012 ACM SIGMOD International Conference on Management of Data},
  pages={229--240},
  year={2012}
}

@inproceedings{napoly_2,
  title={An Overlay Architecture for Pattern Matching},
  author={Karakchi, Rasha and Daniels, Charles and Bakos, Jason},
  booktitle={2019 IEEE 30th International Conference on Application-specific Systems, Architectures and Processors (ASAP)},
  volume={2160},
  pages={165--172},
  year={2019},
  organization={IEEE}
}

@article{streams_on_wires,
  title={Streams on wires: a query compiler for FPGAs},
  author={Mueller, Rene and Teubner, Jens and Alonso, Gustavo},
  journal={Proceedings of the VLDB Endowment},
  volume={2},
  number={1},
  pages={229--240},
  year={2009},
  publisher={VLDB Endowment}
}

@inproceedings{hawk,
  title={Hawk: Hardware support for unstructured log processing},
  author={Tandon, Prateek and Sleiman, Faissal M and Cafarella, Michael J and Wenisch, Thomas F},
  booktitle={2016 IEEE 32nd International Conference on Data Engineering (ICDE)},
  pages={469--480},
  year={2016},
  organization={IEEE}
}

@article{high_frequency_trading,
  title={Efficient event processing through reconfigurable hardware for algorithmic trading},
  author={Sadoghi, Mohammad and Labrecque, Martin and Singh, Harsh and Shum, Warren and Jacobsen, Hans-Arno},
  journal={Proceedings of the VLDB Endowment},
  volume={3},
  number={1-2},
  pages={1525--1528},
  year={2010},
  publisher={VLDB Endowment}
}

@inproceedings{napoly_1,
  title={A dynamically reconfigurable automata processor overlay},
  author={Karakchi, Rasha and Richards, Lothrop O and Bakos, Jason D},
  booktitle={2017 International Conference on Reconfigurable Computing and FPGAs (ReConFig)},
  pages={1--8},
  year={2017},
  organization={IEEE}
}

@ARTICLE{micron_ap,
author={P. {Dlugosch} and D. {Brown} and P. {Glendenning} and M. {Leventhal} and H. {Noyes}},
journal={IEEE Transactions on Parallel and Distributed Systems},
title={An Efficient and Scalable Semiconductor Architecture for Parallel Automata Processing},
year={2014},
volume={25},
number={12},
pages={3088-3098},
doi={10.1109/TPDS.2014.8}
}

@INPROCEEDINGS{bit_split,
author={ {Lin Tan} and T. {Sherwood}},
booktitle={32nd International Symposium on Computer Architecture (ISCA'05)},
title={A high throughput string matching architecture for intrusion detection and prevention},
year={2005},
volume={},
number={},
pages={112-122},
doi={10.1109/ISCA.2005.5}}

@inproceedings{hare,
  title={HARE: Hardware accelerator for regular expressions},
  author={Gogte, Vaibhav and Kolli, Aasheesh and Cafarella, Michael J and D'Antoni, Loris and Wenisch, Thomas F},
  booktitle={2016 49th Annual IEEE/ACM International Symposium on Microarchitecture (MICRO)},
  pages={1--12},
  year={2016},
  organization={IEEE}
}

@mastersthesis{jakob,
author={J. {Meier}},
title={Tools for Cache Coherence Protocol Interoperability},
year={2020},
month=mar,
school={ETH Zürich},
note={Masters thesis, ETH Zürich}
}

@inproceedings{enzian,
  title={Tackling Hardware/Software co-design from a database perspective},
  author={Alonso, Gustavo and Roscoe, Timothy and Cock, David and Owaida, Muhsen and Kara, Kaan and Korolija, Dario and Wang, Zeke and others},
  booktitle={Proceedings of the 6th biennial Conference on Innovative Data Systems Research (CIDR), Amsterdam, Netherlands, January 2020.},
  year={2020}
}

@manual{ultrascale_clb,
title={UltraScale Architecture Configurable Logic Block - UG574},
organization={Xilinx},
address={Xilinx},
year={2017},
note={v1.5},
month=feb,
}

@ARTICLE{coinor,
  author={Lougee-Heimer, R.},
  journal={IBM Journal of Research and Development}, 
  title={The Common Optimization Interface for Operations Research: Promoting open-source software in the operations research community}, 
  year={2003},
  volume={47},
  number={1},
  pages={57-66},
  doi={10.1147/rd.471.0057}
}

@misc{gurobi,  
   author = "Gurobi Optimization, LLC",  
   title = "Gurobi Optimizer Reference Manual",  
   year = 2021,  
   url = "http://www.gurobi.com"
}

@misc{interlaken,
  title={Interlaken Protocol Definition: A Joint Specification of Cortina Systems and Cisco Systems, Rev. 1.2, 52 pp.(Oct. 7, 2008)},
  author={Various},
  howpublished={\url{http://interlakenalliance.com/wp-content/uploads/2019/12/Interlaken_Protocol_Definition_v1.2.pdf}},
  year={2008},
  note={Accessed: 2021-09-02}
}

@Manual{alveou280,
	title = 	 {{Alveo U280 Data Center Accelerator Card Data Sheet}},
	organization = {Xilinx},
	edition = 	 {v.1.3},
	month = 	 may,
	year = 	 2020,
	note = 	 {\url{https://www.xilinx.com/products/boards-and-kits/alveo/u280.html}}}

@Manual{alveou250,
	title = 	 {{Alveo U200 and U250 Data Center Accelerator Cards Data Sheet}},
	organization = {Xilinx},
	edition = 	 {v.1.3.1},
	month = 	 may,
	year = 	 2020,
	note = 	 {\url{https://www.xilinx.com/products/boards-and-kits/alveo/u250.html}}}

@misc{Mellanox:2020,
	author = {Mellanox},
	title = {{Mellanox} {Innova}\texttrademark-{2 FlexOpen} Programmable {SmartNIC}},
	howpublished = {\url{https://www.mellanox.com/files/doc-2020/pb-innova-2-flex.pdf}},
	year = {2020}
}

@book{crockett_zynq_2014,
	title={The Zynq Book: Embedded Processing with the Arm Cortex-A9 on the Xilinx Zynq-7000 All Programmable SoC},
	author={Crockett, Louise Helen and Elliot, Ross and Enderwitz, Martin and Stewart, Robert},
	year={2014},
	publisher={Strathclyde Academic Media}
}

@Misc{ccix,
	author={{CCIX Consortium and others}},
	title={{Cache Coherent Interconnect for Accelerators (CCIX)}},
	howpublished={\url{http://www.ccixconsortium.com}},
	month=jan,
	year=2019
}

@Misc{cxl,
	author = 	 {{CXL Consortium}},
	title = 	 {{Compute Express Link}},
	howpublished = {\url{https://www.computeexpresslink.org/}},
	month = 	 may,
	year = 	 2020}

@article{Choi:2019:IDAM,
	author     = {Choi, Young-Kyu and Cong, Jason and Fang, Zhenman and Hao, Yuchen and Reinman, Glenn and Wei, Peng},
	title      = {{In-Depth Analysis on Microarchitectures of Modern Heterogeneous CPU-FPGA Platforms}},
	year       = {2019},
	issue_date = {April 2018},
	publisher  = {Association for Computing Machinery},
	address    = {New York, NY, USA},
	volume     = {12},
	number     = {1},
	issn       = {1936-7406},
	url        = {https://doi.org/10.1145/3294054},
	doi        = {10.1145/3294054},
	journal    = {ACM Trans. Reconfigurable Technol. Syst.},
	month      = feb,
	articleno  = {4},
	numpages   = {20}
}

@inproceedings{Cock:2022:Enzian,
author = {Cock, David and Ramdas, Abishek and Schwyn, Daniel and Giardino, Michael and Turowski, Adam and He, Zhenhao and Hossle, Nora and Korolija, Dario and
	Licciardello, Melissa and Martsenko, Kristina and Achermann, Reto and Alonso, Gustavo and Roscoe, Timothy},
title = {Enzian: an open, general, {CPU/FPGA} platform for systems software research},
year = {2022},
publisher = {Association for Computing Machinery},
address = {New York, NY, USA},
booktitle = {Proceedings of the 27th ACM International Conference on Architectural Support for Programming Languages and Operating Systems},
pages = {1–18},
numpages = {18},
location = {Lausanne, Switzerland},
series = {ASPLOS 2022}
}

\end{document}